\documentclass[%
 reprint,
 amsmath,amssymb,
 aps,
]{revtex4-2}

\usepackage{graphicx}%
\usepackage{dcolumn}%
\usepackage{booktabs}%
\usepackage{bm}%
\usepackage{hyperref}%
\usepackage{siunitx}%
\usepackage{physics}
\usepackage{xfrac}
\usepackage{csquotes}
\usepackage{listings}
\usepackage{float}

\usepackage{todonotes}

\DeclareMathOperator{\variance}{Var}
\DeclareMathOperator{\covariance}{Cov}

\begin{document}

\preprint{TBD}

\newcommand{\vect}{\bm}
\newcommand{\lik}{\mathcal{L}}
\newcommand{\expect}{\mathbb{E}}

\title{Covering Unknown Correlations in Bayesian Priors by Inflating Uncertainties}

\author{Lukas Koch}
 \email{lukas.koch@uni-mainz.de}
\affiliation{%
 Johannes Gutenberg University Mainz\\
 Institute of Physics - ETAP\\
 Staudingerweg 7\\
 55128 Mainz
}%

\date{\today}%

\begin{abstract}
Bayesian analyses require that all variable model parameters are given a prior probability distribution.
This can pose a challenge for analyses where multiple experiments are combined if these experiments use different parameterisations for their nuisance parameters.
If the parameters in the two models describe exactly the same physics, they should be 100\% correlated in the prior.
If the parameters describe independent physics, they should be uncorrelated.
But if they describe related or overlapping physics, it is not trivial to determine what the joint prior distribution should look like.
Even if the priors for each experiment are well motivated, the unknown correlations between them can have unintended consequences for the posterior probability of the parameters of interest, potentially leading to underestimated uncertainties.
In this paper we show that it is possible to choose a prior parametrisation that ensures conservative posterior uncertainties for the parameters of interest under some very general assumptions.

\end{abstract}

\maketitle

\section{Introduction}

Bayesian analyses require that all variable model parameters are given a prior probability distribution.
This can pose a challenge for analyses where multiple experiments are combined if these experiments use different parameterisations for their nuisance parameters.
If the parameters in the two models describe exactly the same physics, they should be 100\% correlated in the prior.
There is no reasonable way how the physics in one experiment should be different from the physics in the other experiment.
If the parameters describe independent physics, they should be uncorrelated,
unless the external constraints on these unrelated processes are correlated.

But if the nuisance parameters for the two experiments describe related or overlapping physics, it is not trivial to determine what the joint prior distribution should look like.
Imagine, for example, a neutrino oscillation analysis that combines inputs from two long-baseline neutrino beam experiments.
Each experiment has parametrised and quantified uncertainties due to neutrino interaction cross-section model uncertainties, but the parametrisations differ.
One experiment might have chosen to introduce parameters that scale the total cross sections of Charged-Current Quasi-Elastic interactions and Resonant interactions before Final State Interactions (FSI),
while the other decided to introduce a parameters that change the average number of pions and protons that exit the nucleus after FSI.
These two different parameterisations clearly affect the observable data in similar ways, but how exactly a set of values for one should correspond to a set of values (and uncertainties) for the other is not obvious.
The example above is made up, but this kind of problem arose in the context of the joint oscillation analysis of T2K and NOvA\cite{Abubakar2025}.

Even if the priors for each experiment are well motivated, the unknown correlations between them can have unintended consequences for the posterior probability of the parameters of interest, potentially leading to underestimated uncertainties, as we will show in \autoref{sec:problem}.
The authors of the T2K-NOvA paper tackled this issue by explicitly studying the effect of correlations between their most significant nuisance parameters.
They found that the effects are small compared to the overall uncertainties of the result.
This kind of explicit study is labour intensive, so it was only done for a subset of possible combinations of correlations between parameters.
Furthermore, while it is a perfectly valid choice to ignore possible correlations once it is determined that their effect on the analysis would be small,
it would be better to still cover these potential effects.
This would avoid underestimating the total uncertainty due to \enquote{attrition} from many small effects that could add up to a noticeable change in posterior uncertainty,
even if there is no single potential correlation that contributes significantly.

Thankfully, it is possible to inflate the uncorrelated prior covariance in a way that ensures conservative posterior uncertainties for the parameters of interest, as we will see in \autoref{sec:solution}.
This is similar to the prescription in \cite{Koch2025}, where a similar problem of fitting parameters to data sets with missing correlation information is discussed.
This is safe as long as the effect of the nuisance parameters on the posterior distribution of the parameters of interest is linear on the scale of the uncertainties of the nuisance parameters.
In \autoref{sec:higher-order}, we will investigate the consequences of higher order terms.

\section{Potential for underestimated uncertainties}
\label{sec:problem}

It can be easily shown how correlations in the prior assumptions can influence the resulting
posterior uncertainty of parameters of interest.
Let us assume we are dealing with a single parameter of interest $\theta$ and a vector of nuisance parameters $\vect\phi$.
Given some observed data $\vect{x}$ and thus a joint posterior proibability density $f(\theta, \vect{\phi} | \vect{x})$, the posterior variance of $\theta$ can be expressed using the law of total variance\cite[P292]{Soch2025}:
\begin{align}
    \variance[\theta|\vect{x}] &= \expect[\variance[\theta~|~\vect{x},\vect{\phi}]~|~\vect{x}] + \variance[\expect[\theta ~|~ \vect{x},\vect{\phi}]~|~\vect{x}].
\end{align}
So the total variance consists of the expectation value for the conditional variance of $\theta$ given some value for $\vect{\phi}$, plus the variance of the conditional expectation value of $\theta$ given some value for $\vect{\phi}$.
The outer expectation value and variance are both evaluated over the posterior distribution of $\vect{\phi}$, $f(\vect{\phi}|x)$.

Let us assume that the expectation value of the conditional variance of $\theta$ is constant, and does not depend on the particularities of $f(\vect{\phi}|x)$:
\begin{align}
    \expect[\variance[\theta~|~\vect{x},\vect{\phi}]~|~\vect{x}] &= \sigma^2_{\theta|x,\phi}
\end{align}
We can call this the \enquote{intrinsic} variance of $\theta$. Even if we knew the nuisance parameters perfectly well, we could not determine $\theta$ to a higher precision than this.

Let us further assume that we can express the conditional expectation value of $\theta$ as a linear function on the scale of the uncertainties of $\vect{\phi}$:
\begin{align}
    \expect[\theta | \vect{x},\vect{\phi}] &= \theta_0 + \qty(\eval{\vect{\nabla_\phi}\expect[\theta | \vect{x},\vect{\phi}]}_{\vect{\phi}=\vect{\phi}_0})^T \qty(\vect{\phi} - \vect{\phi}_0) \nonumber \\
    &= \theta_0 + \vect{a}^T\vect{\phi} - \vect{a}^T\vect{\phi}_0 = \theta_0' + \vect{a}^T\vect{\phi}.
\end{align}
Here $\vect{a}$ is the gradient of the conditional expectation value of $\theta$ at $\vect{\phi}_0$ and describes which direction in the $\vect{\phi}$ parameter space $\theta$ is sensitive to.
Note that the conditional expectation value is a function of $\vect{\phi}$ that is evaluated for given values of $\vect{\phi}$.
Thus $\vect{a}$ does \emph{not} depend on the distribution of $\vect{\phi}$.

Since we now expressed the conditional expectation value as a linear transform of the nuisance parameters, we can also easily express the variance of it:
\begin{align}
    \variance[\expect[\theta ~|~ \vect{x},\vect{\phi}]~|~\vect{x}] &= \vect{a}^T \Sigma_{\phi|x} \vect{a},
\end{align}
where $\Sigma_{\phi|x}$ is the covariance matrix of the posterior distribution of $\vect{\phi}$.
Let us call this the \enquote{extrinsic} variance of the parameter of interest.

If the nuisance parameters are well constrained by the data, the shape of the priors does not matter.
So for this problem, we can further assume that the posterior covariance matrix of $\vect{\phi}$ is essentially the same as the prior with its $n_B$ blocks of known covariance and unknown correlations between the blocks:
\begin{align}
    \Sigma_{\phi|x} &= \Sigma_{\phi} = \mqty(\dmat[?]{\Sigma_1, \Sigma_2, \ddots, \Sigma_{n_B}}).
\end{align}
This means the total posterior variance of the parameter of interest becomes:
\begin{align}
    \variance[\theta|\vect{x}] &= \sigma^2_{\theta|x,\phi} + \vect{a}^T \Sigma_\phi \vect{a}.
\end{align}
If we assume, for example, two nuisance parameters $\phi_1$ and $\phi_2$,
and that $\theta$ is sensitive to the sum of the two, we get:
\begin{align}
    \vect{a} &= \mqty(1 \\ 1), \\
    \Sigma_\phi &= \mqty(\sigma_1^2 & \rho\sigma_1\sigma_2 \\ \rho\sigma_1\sigma_2 & \sigma_2^2), \\
    \variance[\theta|\vect{x}] &= \sigma^2_{\theta|x,\phi} + \sigma_1^2 + 2\rho\sigma_1\sigma_2 + \sigma_2^2.
\end{align}
Depending on the relative sizes of $\sigma^2_{\theta|x,\phi}$, $\sigma_1$, $\sigma_2$, and $\vect{a}$, the choice of correlation in the prior could have a significant influence on the final reported uncertainty of $\theta$.

\section{Ensuring conservative posteriors}
\label{sec:solution}

Given some pre-determined variances of the nuisance parameters, it is not possible to choose a correlation between them that will always maximise the resulting uncertainty for all parameters of interest.
Whether the posterior uncertainty is increased by correlation or anti-correlation in the the nuisance parameters, depends on the orientation of the gradient $\vect{a}$ in the nuisance parameter space.
In order to ensure a conservative posterior variance, one has to modify the prior variances as well.

Let $\Sigma_{\phi,0}$ be the prior covariance matrix of the nuisance parameters assuming the unknown correlations are all 0.
Further, let $W$ be a suitable block-wise whitening transform of $\Sigma_{\phi,0}$ that acts on the blocks of known covariance:
\begin{align}
    \Sigma_{\phi,0} &= \mqty(\dmat[0]{\Sigma_1,\Sigma_2,\ddots,\Sigma_{n_B}}), \\
    W &= \mqty(\dmat[0]{W_1,W_2,\ddots,W_{n_B}}), \\
    W_i \Sigma_i W_i^T &= I \qquad\forall i,\\
    \Sigma_W &= W \Sigma_\phi W^T,\\
    \vect{b} &= W^{-1T} \vect{a}.
\end{align}
We can then calculate the ratio of the largest possible extrinsic variance to the extrinsic variance we get when assuming no correlations:
\begin{align}
    \alpha &= \frac{\max_{\covariance[\vect{\phi}]} \vect{a}^T \Sigma_{\phi}\vect{a}}{\vect{a}^T \Sigma_{\phi,0}\vect{a}} \nonumber \\
    &= \frac{\max_{\covariance[\vect{\phi}]} \vect{b}^T (W\Sigma_{\phi}W^T)\vect{b}}{\vect{b}^T I \vect{b}} \\
    &\le \max_{\covariance[\vect{\phi}]} \lambda_{W,\text{max}} \\
    &\le n_B \label{eq:limit}
\end{align}
where $\lambda_{W,\text{max}}$ is the largest eigenvalue of the matrix $\Sigma_W$,
and  $n_B$ is the number of blocks of known correlation in $\Sigma_\phi$.

The limit in \autoref{eq:limit} can be seen from the structure of $\Sigma_W$:
\begin{align}
    \Sigma_W &= W\Sigma_{\phi}W^T = \mqty(\dmat[?]{I,I,\ddots,I}).
\end{align}
The blocks on the diagonal are identity matrices by construction.
The off-diagonal blocks are unknown and vary with varying assumptions of the correlations between the blocks in $\Sigma_\phi$.
This means the trace, and thus the sum of eigenvalues of $W\Sigma_{\phi}W^T$ is constant.
This matrix is positive semi-definite, so all eigenvalues are non-negative.
Thus, the only way to increase the value of the largest eigenvalue is to lower other eigenvalues at the same time.
Since all variables in this space are perfectly uncorrelated to the other variables in the same block,
a variable can only be perfectly correlated to as many other variables as there are other blocks.
If it is 100\% correlated to one variable of another block, it cannot be correlated to another variable in the block, since the variables within the block are uncorrelated.
Thus we can only \enquote{concentrate} $n_B$ eigenvalues into a single one.

So, compared to the extrinsic variation assuming no correlations between blocks,
the most conservative choice of correlations for a given gradient vector $\vect{a}$ yields
an extrinsic variance for the parameter of interest at most $n_B$ times larger.
Thus, we can ensure a conservative coverage by simply inflating the uncorrelated prior variance by $n_B$:
\begin{align}
    \Sigma_{\phi,\text{conservative}} = n_B \Sigma_{\phi,0}.
\end{align}

\section{Higher order effects}
\label{sec:higher-order}

We have made a couple of assumptions about the posterior probability distribution to arrive at the conclusion of the previous section:
\begin{enumerate}
    \item The expected variance of the parameters of interest does not depend on the changes to the priors of the nuisance parameters that we are considering.
    \item The expected value of the parameters of interest as a function of the nuisance parameters can be expressed as a linear function on the scale of the nuisance parameter uncertainties.
\end{enumerate}
We will now investigate what happens if we relax these conditions.

If the intrinsic variance of $\theta$ has a linear dependence on $\vect\phi$,
nothing changes, since we are not modifying the expectation value of the prior, only the covariance:
\begin{align}
    \variance[\theta~|~\vect{x},\vect{\phi}] &= \sigma^2_{\theta|x,\phi,0} + \vect{c}^T(\vect\phi-\vect{\phi}_0), \\
    \expect[\variance[\theta~|~\vect{x},\vect{\phi}]~|~\vect{x}] &= \sigma^2_{\theta|x,\phi,0}.
\end{align}
Only quadratic terms or higher will have an effect:
\begin{align}
    \variance[\theta~|~\vect{x},\vect{\phi}] =&~ \sigma^2_{\theta|x,\phi,0} + (\vect{\phi}-\vect{\phi}_0)^T C (\vect{\phi}-\vect{\phi}_0) \nonumber \\
    =&~ \sigma^2_{\theta|x,\phi,0} + \vect{\phi}^T C \vect{\phi} \nonumber \\
    &- 2 \vect{\phi}_0^T C \vect{\phi} + \vect{\phi}_0^T C \vect{\phi}_0 ,\\
    \expect[\variance[\theta~|~\vect{x},\vect{\phi}]~|~\vect{x}] =&~ \sigma^2_{\theta|x,\phi,0} + \Tr[C \Sigma_\phi] \nonumber \\
    =&~ \sigma^2_{\theta|x,\phi,0} \nonumber \\
    &+ \Tr[W^{-1T} C W^{-1} W \Sigma_\phi W^{T}] \nonumber \\
    =&~ \sigma^2_{\theta|x,\phi,0} + \Tr[D \Sigma_{W}].
\end{align}
This is safe as long as $C$ and thus $D$ are symmetric and positive semi-definite.
Because in that case \cite{Zhang2006}:
\begin{align}
    \Tr[D\Sigma_W] &\leq \Tr[D] \lambda_{W,\text{max}} \\
    &\leq \Tr[D] n_B = n_B \Tr[C \Sigma_{\phi,0}], \label{eq:safe}
\end{align}
where the inequality in \autoref{eq:safe} follows the same arguments as before.
The potential contribution by the quadratic term when choosing the \enquote{correct} correlation in the prior is always less than or equal to the quadratic contribution when inflating the uncorrelated prior.

If $C$ is positive semi-definite, that means that the quadratic term in the expansion of $\variance[\theta~|~\vect{x},\vect{\phi}]$ only ever increases the intrinsic variance compared to the linear approximation.
So, by inflating the prior uncertainty, the average intrinsic uncertainty is also increased,
since we are allowing the parameters to deviate further from the mean $\vect{\phi}_0$ and thus increase the contribution of the quadratic terms.
As shown above, this increase is at least as large as the largest possible increase from fine tuning the unknown correlation parameters in $\Sigma_\phi$.

If $C$ is \emph{not} positive semi-definite, that means that the quadratic term in the expansion of $\variance[\theta~|~\vect{x},\vect{\phi}]$ can decrease the intrinsic variance compared to the linear approximation in at least some directions of the parameter space.
In this case, depending on the details of $C$ and $\Sigma_{\phi,0}$, an inflation of the prior variance of $\vect{\phi}$ could even \emph{decrease} the average intrinsic variance of $\theta$.
We have:
\begin{align}
    \Tr[D\Sigma_W] &\geq \Tr[\Sigma_W] \lambda_{D,\text{min}} = n_i \lambda_{D,\text{min}},\\
    \Tr[D\Sigma_W] &\leq \Tr[\Sigma_W] \lambda_{D,\text{max}} = n_i \lambda_{D,\text{max}},\\
    n_B \Tr[C \Sigma_{\phi,0}] &= n_B \Tr[D] \geq n_B  n_i \lambda_{D,\text{min}},
\end{align}
where the inequalities use the fact that $D$ is symmetric and $\Sigma_W$ is positive semi-definite\cite{Zhang2006}.
$\lambda_{D,\text{max}}$ and $\lambda_{D,\text{min}}$ are the largest and smallest (potentially negative) eigenvalues of $D$.
The trace of $\Sigma_W$ is constant, since we are only maximising over the unknown off-diagonal covariances.
Since the diagonal of $\Sigma_W$ is all 1 by construction, the trace is equal to the total number of nuisance parameters $n_i$. 
So the quadratic contribution from the inflation of the uncorrelated prior is again safe if $\lambda_{D,\text{max}} \leq n_B \lambda_{D,\text{min}}$.

If the quadratic contribution to the intrinsic variance is not universally safe,
one can compare the maximum possible contributions $n_i \lambda_{D,\text{max}}$ and $n_i \lambda_{D,\text{min}}$ to the total variance to judge whether this is a problem.

Now we will look at the effect of quadratic terms in the expectation value of $\theta$ as a function of $\vect{\phi}$:
\begin{align}
    \expect[\theta | \vect{x},\vect{\phi}] =&~ \theta_0 + \vect{a}^T (\vect{\phi} - \vect{\phi}_0) \nonumber \\
    &+ (\vect{\phi} - \vect{\phi}_0)^T A (\vect{\phi} - \vect{\phi}_0) \nonumber \\
    =&~ \theta'_0 + (\vect{\phi} - \vect{\phi}'_0)^T A (\vect{\phi} - \vect{\phi}'_0), \\
    \vect{a} =&~ 2A(\vect{\phi}_0 - \vect{\phi}'_0),
\end{align}
with some symmetric matrix $A$.
Assuming a multivariate normal prior distribution of $\vect{\phi}$,
we can calculate the resulting variance:
\begin{align}
    \variance[\expect[\theta ~|~ \vect{x},\vect{\phi}]~|~\vect{x}] =&~ 2\Tr[A\Sigma_\phi A\Sigma_\phi] \nonumber \\
    &+ 4\expect[\vect{\phi} - \vect{\phi}'_0]^T A \Sigma_\phi A \expect[\vect{\phi} - \vect{\phi}'_0] \nonumber \\
    =&~ 2\Tr[A\Sigma_\phi A\Sigma_\phi] \nonumber \\
    &+ 4(\vect{\phi}_0 - \vect{\phi}'_0)^T A^T \Sigma_\phi A (\vect{\phi}_0 - \vect{\phi}'_0) \nonumber \\
    =&~ 2\Tr[A\Sigma_\phi A\Sigma_\phi] + \vect{a}^T \Sigma_\phi \vect{a} \\
    =&~ 2\Tr[B\Sigma_W B\Sigma_W] + \vect{b}^T \Sigma_W \vect{b},
\end{align}
with $B = W^{-1T}AW^{-1}$.
The contribution of the linear term of $\expect[\theta|\vect{x},\vect{\phi}]$ is the same as before, and the trace is the contribution from the quadratic term.

Since $B$ is symmetric, $B^2=B^TB$ is positive semi-definite.
Likewise, since $\Sigma_W$ is positive semi-definite, so is $B^T\Sigma_WB = B\Sigma_WB$.
With this we can constrain the quadratic contribution using similar arguments as before:
\begin{align}
    2\Tr[B\Sigma_W B\Sigma_W] &\leq 2\Tr[B\Sigma_W B] n_B \nonumber \\
    &= 2n_B \Tr[B^2\Sigma_W] \nonumber \\
    &\leq 2n_B^2 \Tr[B^2] \nonumber \\
    &= 2n_B^2 \Tr[A\Sigma_{\phi,0}A\Sigma_{\phi,0}].
\end{align}
So the effect of fine tuning the correlations in $\Sigma_\phi$ will always be less than or equal to an inflation of the prior by $n_B$.
For the extrinsic contribution to the total posterior variance of the parameters of interest, the inflation of priors is safe in the presence of quadratic terms in the expansion of the expected value of $\theta$ as a function of $\vect{\phi}$.

Finally, let us consider the posterior marginal expectation value of $\theta$ using the law of total expectation\cite[P291]{Soch2025}:
\begin{align}
    \expect[\theta | \vect{x}] &= \expect[\expect[\theta ~|~ \vect{x},\vect{\phi}]~|~\vect{x}] = \expect[\mu_{\theta | \vect{x},\vect{\phi}}|\vect{x}].
\end{align}
Like in the previous case of the expectation of the intrinsic variance,
only quadratic terms will have an effect:
\begin{align}
    \expect[\expect[\theta~|~\vect{x},\vect{\phi}]~|~\vect{x}] =&~ \mu_{\theta|x,\phi,0} + \Tr[A \Sigma_\phi] \nonumber \\
    =&~ \mu_{\theta|x,\phi,0} + \Tr[B \Sigma_{W}].
\end{align}
So quadratic terms in the conditional expectation value of $\theta$ move the mean fo the posterior distribution of $\theta$ depending on the assumptions in $\Sigma_\phi$.
There is no \enquote{conservativeness argument} to be made here, since we cannot say which value should be the correct one.
This depends solely on the physics of parameters and what their \enquote{correct} joint distribution should be like.
All we can do is estimate the largest potential mistake we can make when we inflated the prior rather than find the correct correlation from first principles:
\begin{align}
    \Delta\mu_\theta =&~ \max_{\covariance[\phi]} \left|n_B \Tr[B\Sigma_{W,0}] - \Tr[B\Sigma_W]\right| \nonumber \\
    =&~ \max_{\covariance[\phi]} \left|n_B \Tr[B] - \Tr[B\Sigma_W]\right| \nonumber \\
    \leq&~ \max( \left| n_B \Tr[B]- \lambda_{B,\text{max}}\Tr[\Sigma_W] \right|, \nonumber \\
    &\quad \left| n_B \Tr[B] - \lambda_{B,\text{min}}\Tr[\Sigma_W] \right| ) \nonumber \\
    =&~ \max( \left| n_B \Tr[B] - \lambda_{B,\text{max}}n_i \right|, \nonumber \\
    &\quad \left| n_B \Tr[B] - \lambda_{B,\text{min}}n_i \right| ),
\end{align}
where we used the same trace inequalities as before.
$\lambda_{B,\text{max}}$ and $\lambda_{B,\text{min}}$ are the largest and smallest (potentially negative) eigenvalues of $B$.

What values of $\Delta\mu_\theta$ are still acceptable depends on the specific application,
but one general way of interpreting it is to compare it to the square root of the posterior variance of $\theta$.
If the potential bias is much smaller than the posterior uncertainty, it is probably acceptable.

\section{Conclusions}

Doing combined Bayesian analyses of multiple experiments can lead to complications in the selection of priors for nuisance parameters that are different between the experiments, but overlap in their physical meaning.
Depending on the choice of correlation, the resulting posterior variance of parameters of interest can be lower than what it should be.
We have shown that one can ensure a conservative posterior variance of parameters of interest by assuming no correlation between the parameter sets and then inflating the covariance of the nuisance parameter priors by a factor $n_B$, the number of blocks of known, well motivated covariance between parameters, i.e. the number of experiments being combined.

This inflation is always conservative, as long as the effects of the nuisance parameters on the posterior distributions of the parameters of interest can be approximated as linear functions on the scale of the nuisance parameter uncertainties.
Quadratic and higher order terms can potentially lead to lower posterior variances,
but their effect can be estimated as described in \autoref{sec:higher-order}.

Multiplying the variance of the priors of some nuisance parameters by some small integer might not be acceptable in cases where these parameters are a dominating source of uncertainty for the parameter of interest.
In these cases, one will have to deal with this problem with a bespoke solution, like investigating the physics overlap of the two parameter sets in detail and then potentially re-parametrising the uncertainties for a consistent parameter definition between the experiments.
But in the case of sub-dominant parameters, where a doubling or even tripling of the variance has no significant influence on the uncertainty of the parameters of interest,
the inflation of variance as described in this paper is a straight forward way to ensure conservative uncertainties in the posterior distribution.

\section*{Acknowledgments}

I would like to thank Patrick Dunne and Asher Kaboth for discussing with me the joint T2K-NOvA analysis, and the challenges faced therein.
This work was funded by the Deutsche Forschungsgemeinschaft (DFG, German Research Foundation) under Germany’s Excellence Strategy -- EXC 2118 PRISMA+ -- 390831469.

\bibliography{biblio}%

\end{document}